# Achieving higher taxi outflows from a congested drop-off lane: a simulation-based policy study

Fangyi Yang [a, b], Weihua Gu [b], Michael Cassidy [c], Xin Li [b], Tiezhu Li [a*]

[a] *School of Transportation, Southeast University, Nanjing, 210096, China*
[b] *Department of Electrical Engineering, Hong Kong Polytechnic University, Hung Hom, Kowloon, Hong Kong*
[c] *Department of Civil and Environmental Engineering, University of California, Berkeley, USA*

**Abstract**

We examine special lanes used by taxis and other shared-ride services to drop-off patrons at airport and rail terminals. Vehicles are prohibited from overtaking each other within the lane. They must therefore wait in a first-in-first-out (FIFO) queue during busy periods. Patrons are often discharged from vehicles only upon reaching a desired drop-off area near the terminal entrance. When wait times grow long, however, some vehicles discharge their patrons in advance of that desired area. A train station in Eastern China is selected as a case study. Its FIFO drop-off lane is presently managed by policemen who allow taxis to enter the lane in batched fashion. Inefficiencies are observed because curb space near the terminal often goes unused. This is true even when supplemental batches of taxis are released into the lane in efforts to fill those spaces. A microscopic simulation model of a FIFO drop-off lane is developed in-house, and is painstakingly calibrated to data measured at the study site. Simulation experiments indicate that rescinding the FIFO lane's present batching strategy can increase taxi outflow by more than 26%. Further experiments show that even greater gains can be achieved by batching taxis, but requiring them to discharge patrons when forced by downstream queues to stop a prescribed distance in advance of a desired drop-off area. Further gains were predicted by requiring the lead taxi in each batch to discharge its patron(s) only after travelling a prescribed distance beyond a desired location. Practical implications are discussed in light of the present boom in shared-ride services.

## 1. Introduction

Special-use lanes for dropping-off travelers at or near terminals are common features at airports and train stations. Present focus is on fully-separated, single-lane drop-off facilities that are reserved for taxis. These are found in many places in the world, and are especially ubiquitous at high-speed rail terminals throughout China.

Taxis are often made to enter a lane of this sort in batches, meaning in convoys or platoons, and to traverse the single, separated lane in first-in-first-out (FIFO) fashion. The batched outflows achieved in this way depend upon the locations where the lead taxi in each batch stops to discharge its patron(s). Whether or not other taxis discharge their own patrons while their leader dwells curbside will affect outflows too. Whether these other taxis discharge patrons depends upon: a taxi's present position relative to a desired drop-off location; the projected time still required of the taxi to reach that location; and the time already spent to that end.

---

[*] Corresponding author.
Tel: +8613813850110; e-mail: litiezhu@seu.edu.cn.

In short, the operation of FIFO drop-off lanes is rather complex. Importantly, this complex operation can produce sizable delays during busy periods. Patrons tend to be especially sensitive to these delays, since they have planes or trains to catch. Improving the performance of these FIFO lanes thus becomes a worthy objective (e.g., Wu et al., 2018), especially since more of these lanes will be needed for serving taxis and shared-ride vehicles at busy terminals in the future.

Most of the research in this realm pertains to drop-off areas with passing lanes (e.g. Parizi and Braaksma, 1994; Chang et al., 2000; Chang, 2001), and is therefore not of present interest. Some of those studies relied upon deterministic models (Whitlock and Cleary, 1969; Neufville, 1976; Mandle et al., 1980; 1982; Shapiro, 1996; Ashford et al., 2011). The simplicity of these models is desirable, but the inherent variability in vehicle demands, dwell times and drop-off locations are ignored as a result.

Simulation models have often been used to address these variabilities. Yet, the models coded to date overlook other features of FIFO lanes. Some simulation models, for example, ignore that patrons prefer certain drop-off locations over others (Tillis, 1973; McCabe and Carberry, 1975; Hall, 1977). Other logic has failed to appreciate how patrons can grow impatient and opt to alight from taxis in advance of a desired location, particularly when taxi queues grow long (Wang, 1990; Parizi and Braaksma, 1994; Bender and Chang, 1997; Tunasar et al., 1998; Chang et al., 2000; Chang, 2001).

In light of the above, the present work has developed a microscopic simulation model that more faithfully replicates taxi traffic in FIFO drop-off lanes. When in motion, taxi movements are governed by the car-following model in Menendez and Daganzo (2007). Distributions of patrons' desired drop-off locations are estimated from data. So are patron tendencies to grow impatient and alight taxis prior to reaching those desired locations.

The Nanjing South Railway Station (NSR) in Eastern China was selected as a case study. Taxi entries to its FIFO drop-off lane are batched by policemen who are posted at the scene. Once the simulation model was calibrated to replicate observed conditions, it was used to examine alternative schemes for managing the lane's taxi operations. Removing present-day police controls and allowing taxis to enter the lane at will was found to increase taxi discharge flows by 26-32%. Two other control strategies produced even greater gains in outflow. One requires that batched taxis discharge their patrons once stopped by the lead taxi at a location sufficiently close to the desired one. The other requires that a lead taxi discharges its patrons a prescribed distance beyond the desired location, to free-up desirable curb space for other taxis in the batch.

The NSR case-study site is described in the following section. The simulation model's logic is described in section 3. Section 4 presents the data collection process and vehicle trajectory extraction method. Model parameters are estimated in section 5. The model is tested in section 6. Parametric study of the aforementioned control strategies is presented in section 7. Policy implications are discussed in section 8.

## 2. Case Study

The NSR terminal and its taxi drop-off lane are described in section 2.1. Taxi operations in that lane are described in section 2.2.

### 2.1 Site

The drop-off area at the NSR terminal is diagrammed in Figure 1. The area extends for 240m and consists of 5 lanes. Four of the lanes are open to general traffic, and the fifth is reserved for taxis. The taxi lane is separated from others by means of a physical barrier, such that vehicular overtaking is not possible and operation in the lane is FIFO. The figure also shows a painted crosswalk that guides pedestrians to the terminal's entrance and ticketing station. Crossing pedestrians periodically interrupt taxi flows.[1]

Upon alighting from taxis, most patrons proceed directly to the entrance at the center of the drop-off area; i.e. at the 120m mark shown in Figure 1. A numeric minority of patrons proceed instead to ticketing. In an apparent effort to discourage excessive drop-off numbers near the entrance, curbside railing is installed between the 90m and 150m marks.

**2.2 Lane Operations**

During uncongested periods, taxis enter their FIFO lane free of police controls, such that batch size is limited only by the lane's storage space. High taxi demands to the station keep the lane congested for large parts of the day, however.

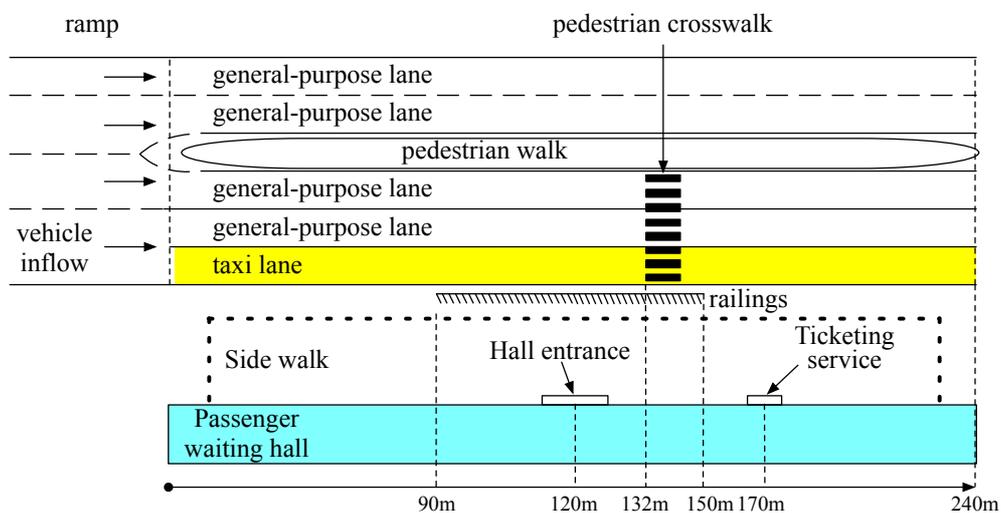

Figure 1. The drop-off area at the NSR Station

During busy periods, a police officer stationed approximately 65m in advance of the drop-off area admits taxis to the FIFO lane in varying-sized batches. Admissions are offered whenever the lane is nearly emptied of its previous batch. The lead taxi in each batch can choose its drop-off location within the 240m area. Data measured from videos show that almost all lead taxis choose locations spanning the 90m and 170m marks. The police officer virtually always releases secondary, smaller-sized batches whenever a present batch is stopped by its leader (stoppages occur whenever leaders discharge their patrons). Admitting secondary batches helps to fill the FIFO lane.

Other taxis in a batch can discharge their patrons while the leader dwells at its drop-off location, or can wait until advancing closer to a desired spot. Taxis that defer discharging their patrons tend to retard outflow from the FIFO lane. Further outflow reductions occur because each fresh batch of taxis is admitted to the lane only after its upstream portion had been empty for some time. Sometimes even the admission of secondary batches leaves upstream space unused. Yet further outflow losses occur because drop-offs almost never occur at the downstream-most portion of the FIFO lane, beyond the 170m mark.

---

[1] A second crosswalk exists downstream of the first (at marker 186m). Pedestrian flows in this downstream crosswalk are small, and seldom interrupt taxi flows.

## 3. Simulation Model

Once in the FIFO lane, taxis move forward as per the car-following model in Menendez and Daganzo (2007). The logic is based on a vehicle's bounded acceleration capabilities, with added considerations for safety and driver/patron comfort. The model was selected for its physical realism and parsimony in parameters, and its details are furnished in the above-cited reference.

Other features of the simulation program are original. These were developed to emulate observations taken at the site, and are described below.

### 3.1 Drop-off decisions

The driver of the lead taxi in a batch chooses a drop-off location as per distributions estimated from data; see section 5.2. The selection process for all other taxis is modelled as per the state transition diagram in Figure 2. A summary is offered below.

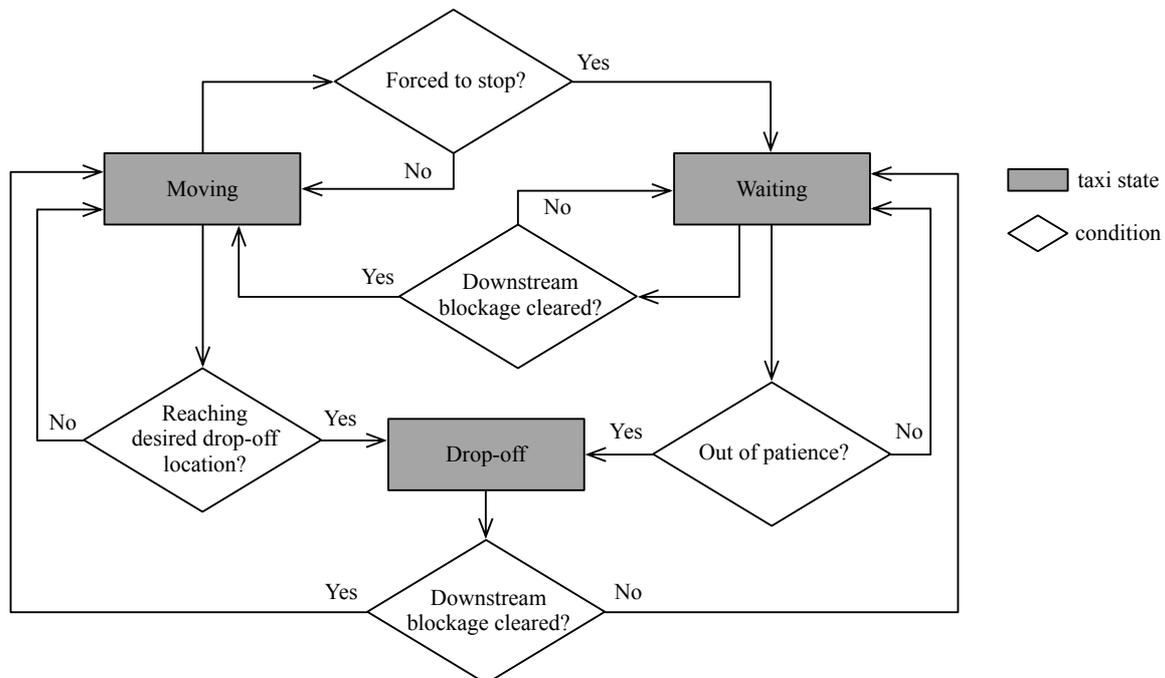

Figure 2. State transition diagram for a taxi that is not the leading one in a batch

Consider a taxi forced by its leader to stop in advance of its desired drop-off location. The taxi drops-off its patron(s) during that stop, if or when its elapsed time at the stop exceeds an underlying limit value (which we term the patron's *patience*). Otherwise, the taxi moves forward once the batch leader enables this. The taxi thereafter discharges its patron: (i) at the first forced stop to occur when the time limit has elapsed; or (ii) upon reaching the desired drop-off location, should that occur first.

Distributions used in executing the above logic were estimated from data, as described in section 5.2. The logic does not consider a taxi's distance from its desired drop-off location when forced to stop. The effect of curbside railing is ignored as well.

## 3.2 Batching Process

To emulate present-day operations at the study site, a taxi batch is admitted to the FIFO lane whenever either: (i) the lane's upstream portion is emptied of taxis over a length of at least $L_{m1}$; or (ii) a length $L_{m2} < L_{m1}$ is cleared of taxis in the lane's upstream portion, and the last taxi in the previous batch continues to dwell in the lane, and has done so for a duration of at least $T_m$. Each secondary batch of taxis is released into the lane using the same logic, but with distinct estimates for parameters $L_{m1}$, $L_{m2}$ and $T_m$.

We further denote $L_{left}$ as the lane space left upstream of a batch when its last taxi is stopped. The number of taxis in a batch is thus determined by dividing the occupied space (i.e., the available space minus $L_{left}$) by the jam, or stopped-vehicle spacing.

## 4. Data Collection and Taxi Trajectory Extraction

Most of the information required in the above models can be derived from vehicle trajectories, which were extracted from videos of vehicle traffic recorded on the site. The data collection process is described in section 4.1. The vehicle trajectory extraction method is presented in section 4.2.

### 4.1 Data collection

We use four video cameras installed at the 20m, 65m, 86m, and 152m marks, as shown in Figure 3. These cameras were placed 5m above the road surface using telescopic poles. Each camera only captures vehicle motions with sufficient resolution for a range of 20-90m ahead of the camera's location. Cameras 1, 3, and 4 were facing downstream of the FIFO lane, while Camera 2 was facing upstream (see the thick arrows in Figure 3), so as to capture vehicle movements in the upstream part of that lane. All four cameras collectively cover the entire 240m lane. Their fields of view are illustrated in Figures 4a-d, respectively.

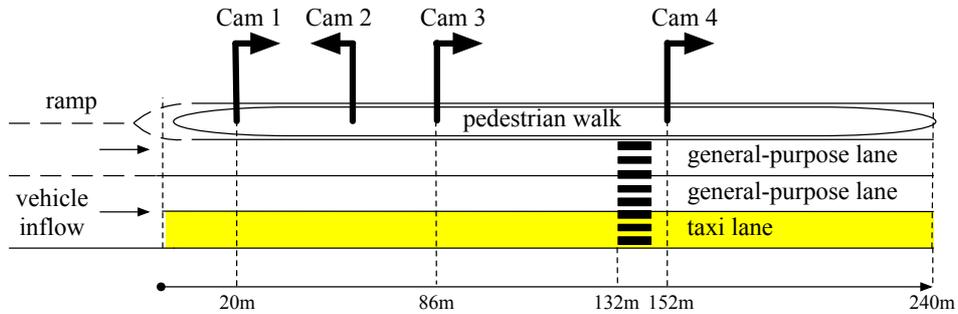

Figure 3. Positions of video cameras

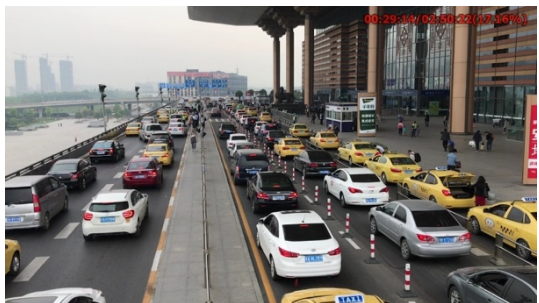

(a) Camera 1

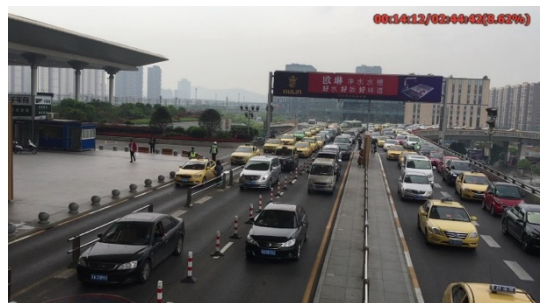

(b) Camera 2

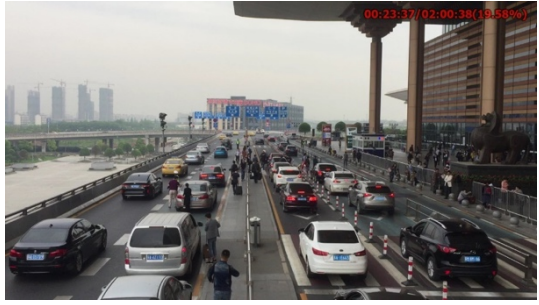 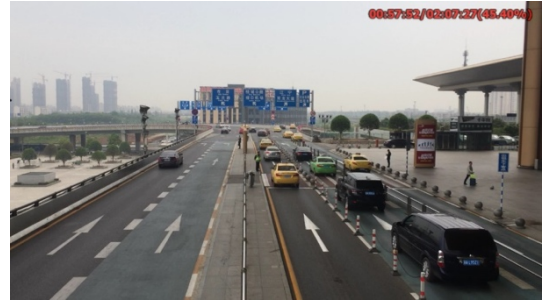

(c) Camera 3　　　　　　　　　　　(d) Camera 4

Figure 4. Fields of view of the four cameras

We filmed two videos in the morning peak periods on April 25 and July 13, 2017, respectively. The April 25 video covers a duration of 1h 33min, from 8:55:50 AM to 10:28:57 AM. Totally 561 vehicles were recorded passing through the FIFO lane. The July 13 video covers a duration of 1h 22min, from 7:59:06 AM to 9:21:35 AM. Totally 465 cars were captured in the lane. Long vehicle queues persisted on both days. Batching control was imposed by police throughout in both videos.

### 4.2 Taxi trajectory extraction

We developed a computer program for generating vehicle trajectories. Via this program, we manually click each vehicle's rear-end location on the video screen for every second, and the program will automatically calculate the longitudinal coordinate (i.e., the mark between 0-240m) of each click in the FIFO lane. The calculation method is relegated to Appendix A. Trajectories are then plotted by connecting these points in a time-space diagram. We further mark the door opening and closing times for each drop-off maneuver. Figure 5 illustrates the extracted vehicle trajectories for time 0:00:00 to 0:18:36 of the video taken on April 25, 2017, where the red horizontal segments indicate drop-off locations and the durations between door opening and closing.

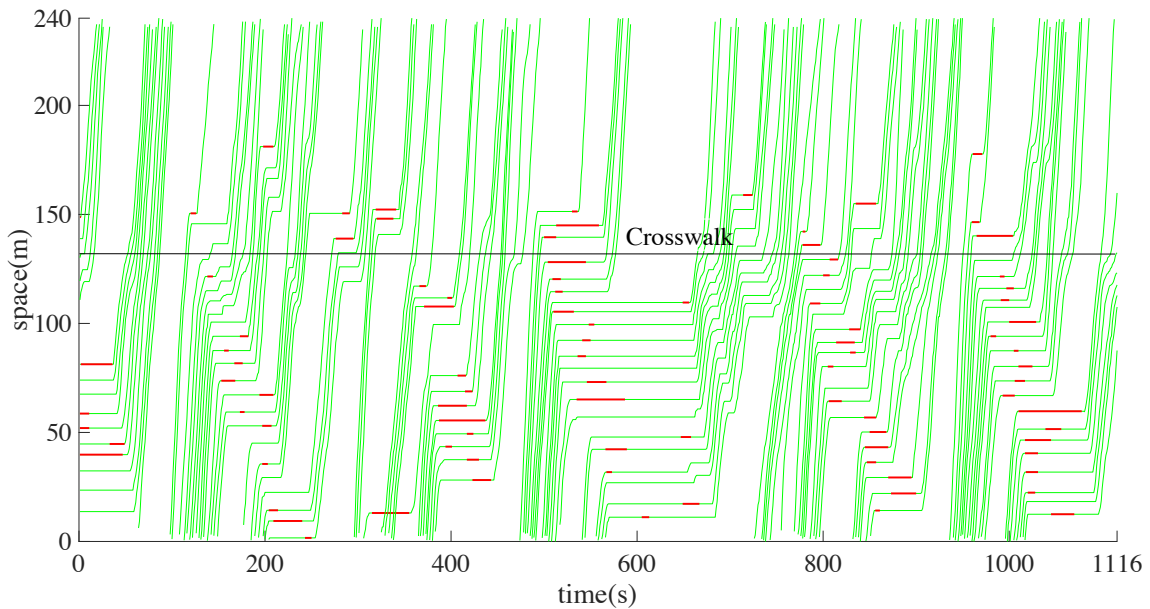

Figure 5. Illustration of vehicle trajectories

## 5. Parameter Estimation

The taxi trajectories were used to estimate parameters in our simulation model. Estimations were performed in three parts. First, the FIFO lane was partitioned into multiple, contiguous segments. This was done because taxis' *forced wait times* varied by location within the lane. The forced wait time is defined for each of its forced stops as: (i) the duration from when the taxi is forced to stop until it starts to drop-off its patron(s); or (ii) the entire duration of that forced stop, should patron drop-off not occur. Partitioning was based on measured patterns in the data, and is distinct across both observation days.

Second, distributions of patrons' patience were estimated for each segment and day; see again section 3.1 for the definition of patience. Note that patience is a latent variable while the forced wait times of taxis are observable. A taxi's forced wait time is bounded by its patience. The third part of the process entailed straightforward estimations of the other parameters used in our simulation model.

The three parts are described below. Certain details are relegated to appendices.

### 5.1 Lane partitioning

The taxis' forced wait times tended to be larger when they occurred toward the upstream end of the FIFO lane. This can be seen in Figures 6a and b. Data shown in figure 6a are those collected on April 25, and correspond to taxis' first instances of being stopped in the lane. (Taxis' second and later instances of forced stops were less common and shorter in duration.) The thin line is a trend line through the data. Note how it tends to move downward with location in the lane. (The trend is not strictly downward for reasons to be explained momentarily.) A similar pattern is also observed in Figure 6b for the forced wait times of first-instance forced stops on July 13.

A $k$-means clustering algorithm (Hartigan and Wong, 1979) was used to partition the FIFO lane into four segments, also as shown in Figure 6 and detailed in Appendix B. Outcomes of segment-specific analyses are presented in Table 1 for each observation day. Note how the lane was partitioned into four segments each day, while the physical lengths of those segments, and the average forced wait times in them varied across days.

Further note how the average forced wait times in Segment 3 were always greater than those in Segment 2 just upstream. This is because the lane's curbside railing coincided with Segment 3 (see again Figure 1), which discouraged patrons from alighting there.

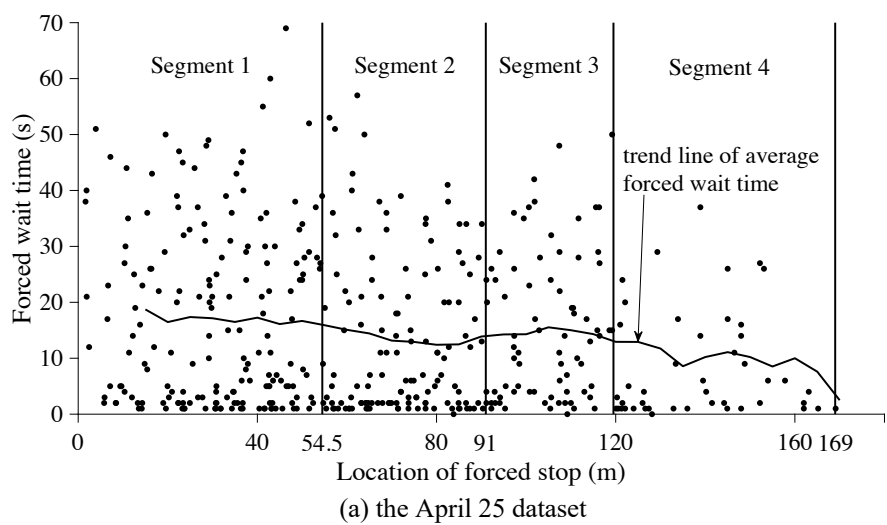

(a) the April 25 dataset

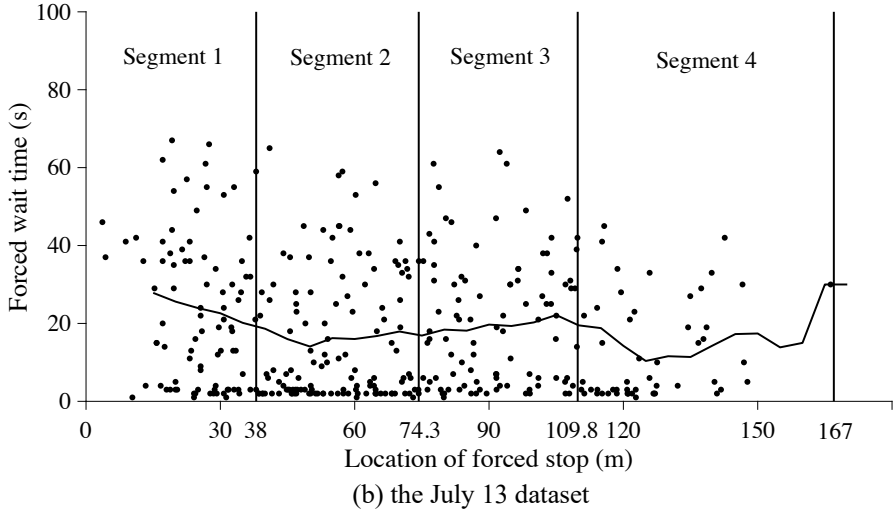

(b) the July 13 dataset

Figure 6. Lane partition ($k = 4$)

Table 1. Four segments and average forced wait times in each

|  |  | **Segment 1** | **Segment 2** | **Segment 3** | **Segment 4** |
|---|---|---|---|---|---|
| April 25 | Range (m) | 0-54.5 | 54.5-91 | 91-119.5 | 119.5-169 |
|  | Average forced wait time (s) | 17.5 | 13.2 | 15.5 | 8.4 |
| July 13 | Range (m) | 0-38 | 38-74.3 | 74.3-109.8 | 109.8-167 |
|  | Average forced wait time (s) | 23.4 | 14.9 | 20.8 | 12.7 |

### 5.2 Distributions of patron patience

For first instances of forced stops, distributions of patron patience were separately estimated for each segment and day. Due to the small number of second, third and fourth instances of forced stops, these were combined and collectively estimated across all segments, but again for each day.

In all cases, the patience durations seem to follow a mixed-distribution pattern, as do the observed forced wait times. (Note that the two are highly correlated.) This is exemplified by the dark bars in Figure 7a. The data in that figure are of first-instance stops in the upstream-most segment on April 25. Note the large number of forced wait times that are close to zero. These denote instances of impatient patrons who alight their taxis soon after stopping. The remaining observations in the figure follow a more scattered distribution.

Patience was therefore described by a mixture of two gamma distributions. Values for their coefficients were obtained via maximum likelihood estimation (MLE). The process is described in Appendix C. The resulting cumulative distribution function (CDF) for the sample data in Figure 7a is shown by the dashed line in Figure 7b. Note in the latter figure how the estimated CDF closely matches the empirical distribution shown by the solid curve.

### 5.3 Remaining parameters

This section estimates other model parameters, including: maximum acceleration and deceleration, desired speed, initial vehicle speeds and headways, drop-off distribution, drop-off duration distributions, parameters for emulating the effect of the pedestrian crosswalk, percentage of vehicles having drop-off requests, and the batching control parameters. The derivations of these parameters are explained one by one in what follows. Tables 2 and 3 summarize the parameter estimates for the April 25 dataset and the July 13 dataset, respectively.

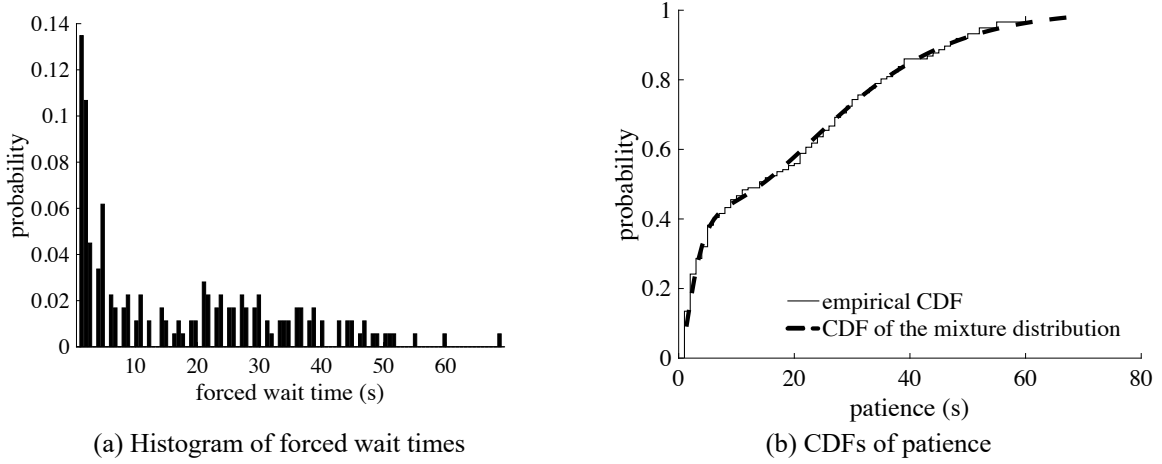

(a) Histogram of forced wait times   (b) CDFs of patience

Figure 7. Patrons' patience distribution for first-instance forced stops in Segment 1 (the April 25 dataset)

**Maximum acceleration and deceleration**. To reduce the errors brought by our manual mouse-click method, we calculate the maximum acceleration for each taxi as follows. First remove the parts of trajectory where the taxi was stopped. From the remaining trajectory, calculate the maximum of the *average* acceleration for every 5-second period in which the taxi was constantly speeding up. The maximum acceleration is then set to the average of the above values across all the taxis. The maximum deceleration is computed in a similar fashion.

**Desired speed**. This parameter was estimated separately for each segment of the FIFO lane, since taxis exhibited different cruise speeds in different segments. This is partly because many taxis kept traveling at a relatively low speed before dropping off passengers, even if the spacing ahead allowed them to accelerate. For each segment, the desired speed was estimated as the maximum average speed over a 7-second period.

**Initial vehicle speed and headway**. We assume all the vehicles enter the drop-off area at a constant initial speed, and with a constant, sufficiently small headway (2 seconds). The initial speed is estimated as the average initial speed measured from the videos.

**Probability that a vehicle makes a drop-off in the FIFO lane**. This parameter was estimated directly from the datasets.

**Distribution of desired drop-off location**. We fit a nonparametric distribution to the drop-off locations of lead taxis of each taxi batch. We assume that desired drop-off locations of all the taxis (no matter whether it is a lead one or not) follow that same distribution.

**Distribution of drop-off duration**. For lead taxis of each batch, a non-parametric distribution of drop-off duration was fit to their dwell times. For taxis that are not lead taxis, another non-parametric distribution was fit to the durations between door opening and closing for each of those taxis. Note that these latter durations do not include: (i) the preparation time before door opening, during which the patrons may unfasten the seat belt and make the payment; and (ii) the post-drop-off time after door closing, during which the driver may store the cash payment and make records. Thus, a fixed time, which is equal to the mean of the sum of items (i) and (ii) above for lead taxis, is added to the non-parametric drop-off duration distribution estimated for non-lead taxis.

Table 2. Other parameter values for the April 25 dataset

| Parameter | | Value |
|---|---|---|
| *Car-following model* | | |
| Reaction time[2] | | 1 s |
| Jam spacing | | 7.5 m |
| Maximum acceleration | | 2.12 m/s$^2$ |
| Maximum deceleration | | -2.86 m/s$^2$ |
| Cruise speed in Segment 1 | | 6.13 m/s |
| Cruise speed in Segment 2 | | 4.94 m/s |
| Cruise speed in Segment 3 | | 3.30 m/s |
| Cruise speed in Segment 4 | | 6.07 m/s |
| Cruise speed in [169,240] m | | 5.72 m/s |
| Initial speed | | 4.54 m/s |
| Initial headway | | 2 s[3] |
| *Drop-off behavior* | | |
| Desired drop-off location distribution | | A nonparametric distribution fitted by data |
| Drop-off duration distribution for lead taxis | | A nonparametric distribution fitted by data |
| Drop-off duration distribution for other taxis | | A nonparametric distribution fitted by data |
| Proportion of taxis that drop-off patrons in the lane | | 0.83 |
| *Signal representing the crosswalk* | | |
| Red period distribution | | A nonparametric distribution fitted by data |
| Green period distribution | | A nonparametric distribution fitted by data |
| *Batching control* | | |
| $L_{m1}$ | Primary batches | 122.4 m |
| | Secondary batches | 70.2 m |
| $L_{m2}$ | Primary batches | 66.5 m |
| | Secondary batches | 15.4 m |
| $T_m$ | Primary batches | 120 s |
| | Secondary batches | 14.5 s |
| $L_{left}$ | Primary batches | 42.3 m |
| | Secondary batches | 12.3 m |

**Pedestrian crosswalk.** Pedestrian interruptions to taxi flows were modelled by assuming the presence of a demand-responsive traffic signal at the crosswalk; see again Figure 1. Red and green phases varied across cycles and mimicked periods when taxi movements were and were not interrupted by pedestrian crossings. Non-parametric distributions were fit to the data for each type of phases.

**Batching control parameters**. Parameters associated with police batching, $L_{m1}$, $L_{m2}$ and $T_m$ (see section 3.2) were estimated for each day's data via the *k*-means clustering algorithm (Hartigan and Wong, 1979). Parameter $L_{left}$ was separately set to the average empty lane space upstream of the primary and secondary batches, again for each day's data.

---

[2] This is a constant that represents the time needed for the backward shockwave to propagate across one vehicle in queue (Daganzo, 2006; Menendez and Daganzo, 2007)
[3] The initial speed and headway were set to ensure that taxis entered the FIFO lane in batches and the safety constraint of car-following model was satisfied. The same values were used for the July 13 dataset.

Table 3. Other parameter values for the July 13 dataset

| Parameter | Value |
|---|---|
| *Car-following model* | |
| Reaction time | 1 s |
| Jam spacing | 7.5 m |
| Maximum acceleration | 2.39 m/s$^2$ |
| Maximum deceleration | -2.37 m/s$^2$ |
| Cruise speed in Segment 1 | 5.91 m/s |
| Cruise speed in Segment 2 | 5.11 m/s |
| Cruise speed in Segment 3 | 4.20 m/s |
| Cruise speed in Segment 4 | 5.89 m/s |
| Cruise speed in [169,240] m | 5.82 m/s |
| Initial speed | 4.48 m/s |
| Initial headway | 2 s |
| *Drop-off behavior* | |
| Desired drop-off location distribution | A nonparametric distribution fitted by data |
| Drop-off duration distribution for lead taxis | A nonparametric distribution fitted by data |
| Drop-off duration distribution for other taxis | A nonparametric distribution fitted by data |
| Proportion of taxis that drop-off patrons in the lane | 0.85 |
| *Signal representing the crosswalk* | |
| Red period distribution | A nonparametric distribution fitted by data |
| Green period distribution | A nonparametric distribution fitted by data |
| *Batching control* | |
| $L_{m1}$ Primary batches | 199.4 m |
| $L_{m1}$ Secondary batches | 70.4 m |
| $L_{m2}$ Primary batches | 72.2 m |
| $L_{m2}$ Secondary batches | 27.7 m |
| $T_m$ Primary batches | 93.5 s |
| $T_m$ Secondary batches | 16.9 s |
| $L_{left}$ Primary batches | 53.4 m |
| $L_{left}$ Secondary batches | 18.3 m |

## 6. Model Testing

Having estimated its input parameters, the model was tested against each day's data measured in the FIFO lane. Each day's predictions were compared against the same data used to calibrate the model.[4] Four metrics were used for comparison. These were the taxi: (i) outflows; (ii) travel times; (iii) forced wait times and their locations; and (iv) drop-off locations. Large demands for the lane (400 taxis/h) were used each day. This ensured that taxi queues persisted in advance of the lane, as per observations. To ensure convergence for the averages of simulated outcomes, all simulated outcomes reported below are averages of 500 runs. Figures 8a and b illustrate the convergence of the mean and standard deviation of taxi travel times, respectively, over 500 simulation runs. Convergence of other metrics is similar and is omitted in the interest of brevity.

---

[4] Our intent is not to demonstrate the generalizability of the simulation model, but to show only that it can faithfully emulate each day's operations under the current batching strategy. These outcomes will serve as baselines against which alternative strategies will be judged.

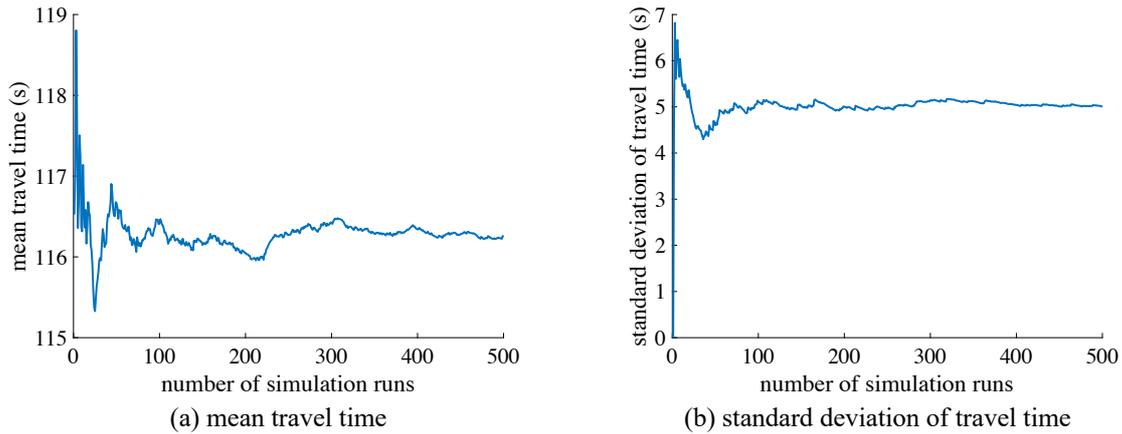

(a) mean travel time  (b) standard deviation of travel time

Figure 8. Convergency of the mean and standard deviation of taxi travel times (the April 25 dataset)

### 6.1 Taxi outflows and travel times

Measured and simulated outflows and travel times are presented in Table 4. Outcomes for both days are shown. Simulated outcomes agree with measured ones to within 6%. The model under-predicted both outflows and travel times. The under-predictions of outflows are partly due to the model's use of a demand-responsive traffic signal to describe how pedestrian crossings interrupt taxi flows; see again section 5.3. In reality, some taxi drivers squeezed their vehicles between neighboring groups of crossing pedestrians. This aggressive behavior slightly increased taxi outflows. Further differences in travel time are likely due to certain unique features of taxi operations that are not captured in the car-following model of Menendez and Daganzo (2007). For example, a taxi driver's search for a suitable drop-off spot might occasionally lead to lower accelerations.

Table 4. Measured and simulated outflows and travel times

| Parameter | Dataset | Type | Value | Relative error |
|---|---|---|---|---|
| Outflow (taxis/h) | April 25 | field data | 361.5 | -5.81% |
| | | simulation | 340.5 | |
| | July 13 | field data | 338.3 | -2.16% |
| | | simulation | 331 | |
| Average travel time (s) | April 25 | field data | 122.8 | -5.62% |
| | | simulation | 115.9 | |
| | July 13 | field data | 106.2 | -0.19% |
| | | simulation | 106 | |

As a further example of the good match between simulated and observed values, Figures 9a and b present each day's measured and simulated probability density functions (PDFs) of travel time. The agreement between measurement and prediction is quite good.

### 6.2 Forced Waits

The occurrences of forced stops and the average forced wait times are shown in Tables 5 and 6 for the two observation days. Note how for each day the total number of forced stops predicted by the simulation model matches the observed numbers to within 7%. Differences in average forced wait times across all stops are larger, but are still less than 10%. The tables also show that the simulation model over-predicted the number of first-instance forced stops occurring in upstream-most Segment 1.

It thus under-predicted the numbers of first-instance stops in nearly all downstream segments.[5] Since many (simulated) taxis chose not to drop off patrons in Segment 1, more second-, third-, and fourth-instance stops are observed in the simulation.

Differences shown in Table 5 and 6 are again linked to the use of the demand-responsive signal for approximating the effect of pedestrian crossings on taxi flows. The aggressive behavior of those real drivers who squeezed their taxis through gaps in pedestrian flows diminished the number of forced stops that actually occurred in Segment 1.

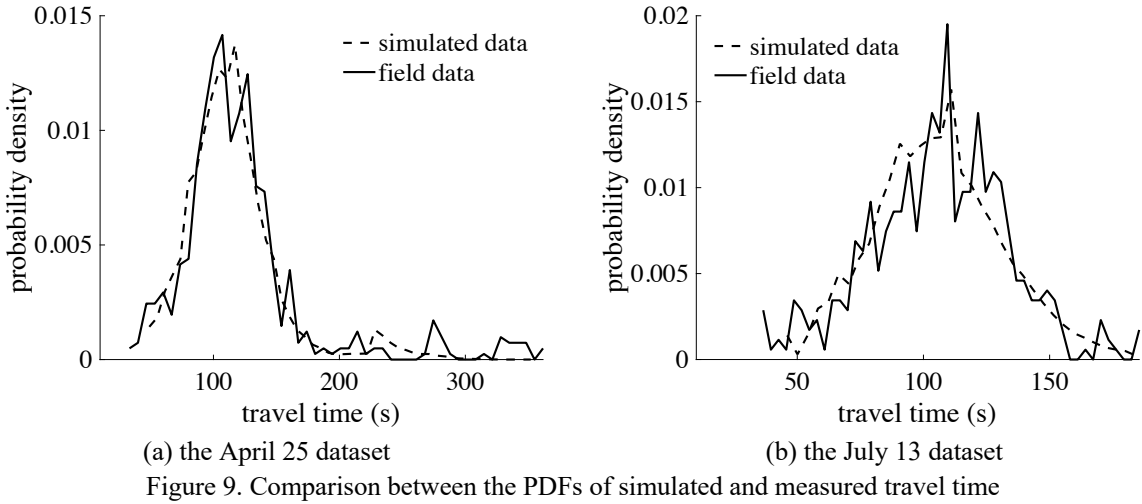

(a) the April 25 dataset  (b) the July 13 dataset
Figure 9. Comparison between the PDFs of simulated and measured travel time

Table 5. Measured and simulated numbers of forced stops and mean forced wait times (the April 25 dataset)

|  |  | 1st-instance forced stops | | | | 2nd~4th-instance forced stops | Total |
|---|---|---|---|---|---|---|---|
|  |  | Zone 1 | Zone 2 | Zone 3 | Zone 4 | | |
| Number of forced stops | field data | 183 | 135 | 79 | 47 | 39 | 483 |
|  | simulation | 198 | 128 | 41 | 22 | 64 | 453 |
| Average forced wait time (s) | field data | 17.5 | 13.2 | 15.5 | 8.4 | 12.2 | 16.9 |
|  | simulation | 17.9 | 15.1 | 16.2 | 11 | 16.8 | 16.5 |

Table 6. Measured and simulated numbers of forced stops and mean forced wait times (the July 13 dataset)

|  |  | 1st-instance forced stops | | | | 2nd~4th-instance forced stops | Total |
|---|---|---|---|---|---|---|---|
|  |  | Zone 1 | Zone 2 | Zone 3 | Zone 4 | | |
| Number of forced stops | field data | 87 | 126 | 96 | 53 | 34 | 396 |
|  | simulation | 100 | 149 | 60 | 23 | 80 | 412 |
| Average forced wait time (s) | field data | 23.4 | 14.9 | 20.8 | 12.7 | 11.5 | 17.7 |
|  | simulation | 20.9 | 14 | 16.1 | 11.6 | 14.7 | 16 |

### 6.3 Drop-off locations

Mean drop-off locations are shown for each day in Table 7. The difference between simulated and measured values is negligible for the April 25 dataset, and falls within 3% for the July 13 data. The simulated mean is slightly smaller than the measured one for the latter dataset. This too was likely caused by having over-predicted the occurrences of forced stops in Segment 1; i.e. more stops predicted in the upstream segment resulted in more drop-offs there as well. This diminished the predicted mean in Table 7.

---

[5] An exception occurred for the July 13 dataset, in which the simulation model over-predicted the number of first-instance forced stops in Segment 2. This occurred because Segments 1 and 2 as partitioned for the July 13 dataset are both far upstream of a desired drop-off location near the station entrance; see again Figure 1 and Table 1.

Table 7. Measured and simulated mean drop-off locations

| Parameter | Dataset | Type | Value |
|---|---|---|---|
| Average drop-off location (m) | April 25 | field data | 79.7 |
| | | simulation | 79.5 |
| | July 13 | field data | 82 |
| | | simulation | 83.8 |

The PDFs for each day's drop-off location are shown in Figures 10a and b. Agreements between measured and simulated values are again quite good.

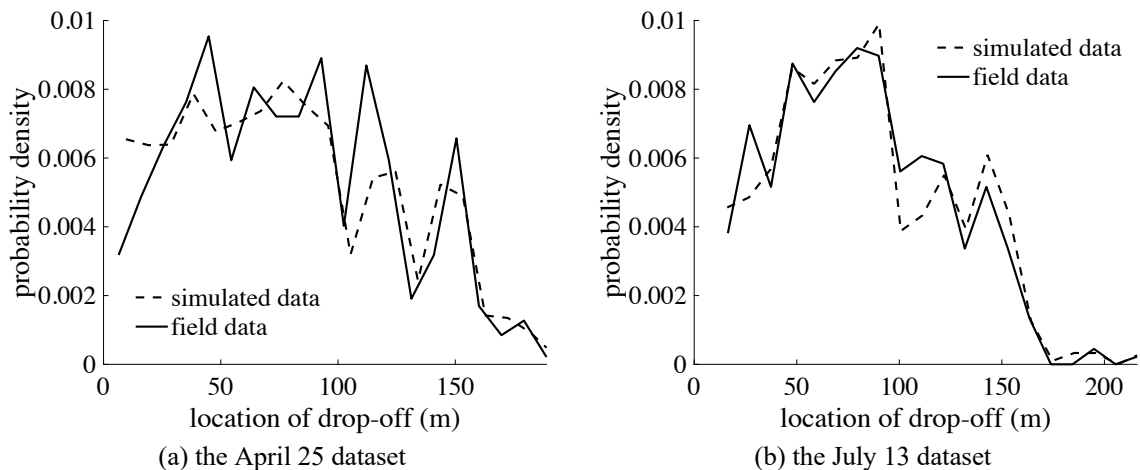

(a) the April 25 dataset    (b) the July 13 dataset

Figure 10. PDFs of taxi drop-off locations

## 7. Experiments

The simulation model is next used to evaluate three alternative strategies for managing taxi movements in the FIFO lane. In all three cases, present-day batching operations (in advance of the FIFO lane) are rescinded. Taxis are free instead to form queues at the entry to the FIFO lane. The first two of these alternatives seek more effective utilization of the FIFO lane's upstream portions. The third seeks this as well, and also enhances utilization of the lane's downstream portions.

The first management strategy is a "no-control" alternative. Once queued taxis enter the lane, they can drop-off patrons there wherever they wish.

The second is a "no-wait" alternative. Upon traveling the lane for a distance of at least $L_0$, taxis must thereafter discharge patrons when first forced by downstream conditions to stop. Should no forced stops occur, a taxi may discharge its patron(s) at any location desired within the lane.

The third alternative also requires taxis to drop-off patrons upon its first force stop beyond a distance $L_0$. When no such stops occur, however, this last alternative requires that taxis discharge patrons only upon reaching a specified distance, $L_H$, toward the FIFO lane's downstream end.

Taxi outflows under each alternative are compared against those simulated under present-day batching. All else equal, higher outflow means less queueing and delay in and around the lane. Outcomes from each comparison are presented in the three subsections to follow. Taxi demand was again set high (700 taxis/h) to ensure persistent queueing at the lane's entry; i.e., to ensure that all outflows reflect maximum (capacity) rates. The simulated outcomes presented below are again averages of 500 simulation runs to ensure that those averages converge.

### 7.1 "No-control" strategy

Outcomes from the first set of comparisons are presented in Table 8. Note how each day's outflow in the absence of control exceeds present-day (simulated) rates by 26-32%. The no-control alternative seems clearly to do a better job of utilizing the FIFO lane's upstream portion: the table shows that average drop-off location under the alternative moves upstream by 13-17m. After removing the batching control, the outflow on July 13 becomes greater than that on April 25 (note that the present-day outflow on July 13 is lower than that on April 25; see Table 4). This is mainly because on July 13 more vacant lane space was left behind each batch of taxis; see the values of $L_{left}$ in Tables 2 and 3.

Table 8. Comparison of simulated outcomes between the present-day control and the no-control alternative

| Parameter | Dataset | Type | Value |
|---|---|---|---|
| Outflow (taxis/h) | April 25 | present-day | 340.5 |
| | | no control | 429.3 |
| | July 13 | present-day | 331 |
| | | no control | 438.4 |
| Average drop-off location (m) | April 25 | present-day | 79.5 |
| | | no control | 66.2 |
| | July 13 | present-day | 83.8 |
| | | no control | 67.2 |
| Number of forced stops in Segment 1 | April 25 | present-day | 198 |
| | | no control | 277 |
| | July 13 | present-day | 100 |
| | | no control | 193 |

### 7.2 "No-wait" strategy

Minimum distance location for drop-offs, $L_0$, was examined parametrically. The full length of the FIFO lane was considered, such that $0 \leq L_0 \leq 240$ m. Setting $L_0$ to the full length of the lane (240m), in effect, allows taxis to drop-off patrons wherever they wish. It is therefore akin to the no-control alternative.

Each day's taxi outflow is plotted in Figure 11 as a function of $L_0$. Note how outflow is maximum when $L_0 = 0$. That choice of $L_0$ makes best use of the lane's upstream segments.

Two related points emerge from the figure as well. First, setting $L_0 = 0$ increases taxi outflow by more than 20% over the no-control alternative (with $L_0 = 240$ m), or by more than 50% over the present-day batching control. Second, benefits of this second alternative disappear when $L_0$ grows sufficiently large. In the present case, the curves trend horizontal when $L_0 > 120$ m.

Of course, setting $L_0$ at a small value may be objectionable to some patrons who find themselves walking long distances from their taxis to the transport terminal. This matter will be taken up in section 8. In the meantime, we turn our attention to the third and final management alternative.

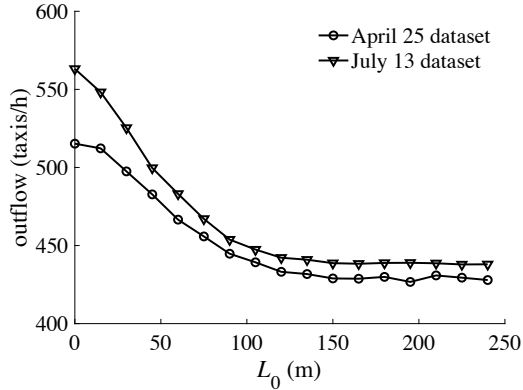

Figure 11. Effect of no-wait policy on the FIFO lane's outflow

### 7.3 Promoting downstream drop-offs

The second, "no-wait" strategy with parameter $L_0$ continues to be in force. Additionally, every taxi not encountering a forced stop downstream of $L_0$ can now drop-off patrons only upon reaching a location $L_H$.

Outflows for $L_H \in [120, 210]$ m are plotted in Figures 12a and b. Recall from Figure 1 that 120m coincides with the entrance to the NSR station. The upper bound of $L_H$ was set to 210m to reflect an assumption that no patron would be willing to alight her taxi beyond this distance. (No taxi was observed to drop-off patrons downstream to the 210m marker.) Separate curves are shown in the figures for $L_0 = 0$, 30, 60 and 90m.[6]

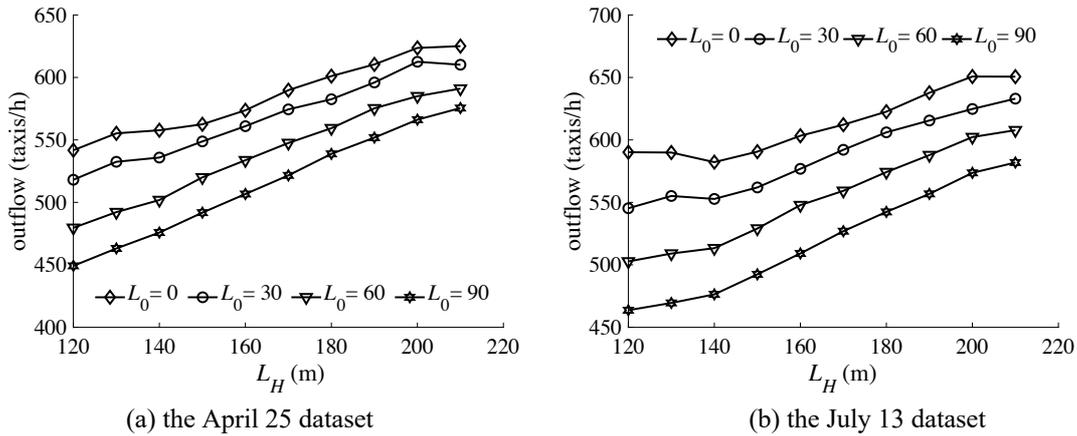

(a) the April 25 dataset  (b) the July 13 dataset

Figure 12. Effect of promoting downstream drop-offs

The figures confirm that outflows increase with smaller $L_0$; i.e. a small $L_0$ leads to better utilization of the lane's upstream segments. Outflows are also shown to increase with large $L_H$, no doubt by promoting better use of the lane's downstream segments. Thus, for example, study of Figures 12a and b shows that raising $L_H$ from 120m to 210m typically increases taxi outflow by about 10-28%.

---

[6] Larger values of $L_0$ were not tested in light of the findings in section 7.2.

Greatest outflows were therefore achieved by $L_0 = 0$ and $L_H = 210$ m. Outflows in this extreme case were more than 83% higher than what is presently achieved via batching. (This becomes evident by comparing outcomes in Table 4 with those in Figures 16a and b.) These extremal thresholds, moreover, increase outflows by over 45% compared to the no-control alternative. And less restrictive thresholds of $L_0 = 90$ m and $L_H = 160$ m still enhance taxi outflows; e.g. by over 49% compared to present-day rates.

Taxi patrons might, of course, object to high values of $L_H$, as well as to low values of $L_0$. Matters of this kind are discussed next.

## 8. Conclusions

Simulations of a busy FIFO drop-off lane unveil the value of managing taxi operations in efficient fashion. The simulation model itself was developed in-house to emulate taxi movements in the lane. Parameters were estimated from data measured over two days. Once separately calibrated to each day's data, the model replicated the day's movements quite well. Outputs were thus used as baselines against which alternative lane-management strategies were compared.

Comparisons show that rescinding the present-day batching strategy can increase taxi outflows from the FIFO lane, and thus diminish delays and queueing. Instituting a "no-control" alternative alone increased outflows by more than 26%. A distinct batching alternative that requires drop-offs whenever downstream conditions force a taxi to stop a distance greater than $L_0$ inside the lane was tested as well. By promoting greater use of curb-space in upstream portions of the FIFO lane, this alternative improved taxi outflows by an additional 20% or more. Coupling this with another requirement that non-stopped taxis discharge patrons at a lengthy distance $L_H$ inside the lane promotes greater use of downstream curb-space. Instituting requirements in terms of both $L_0$ and $L_H$ thus further improved outflows by as much as 20%.

The above predictions are compelling, but are not without errors. The model's failure to consider a patron's accrued delay in choosing her drop-off location is a likely source of error. Further sources may stem from unique features of taxi motion (e.g. in search of drop-off spots), which are not captured in the car-following model selected for the present work. Such is the nature of simulation. Our inability to calibrate a single model to replicate operations in any given day may be a further concern, though in fairness the data suggest that taxi outflows are influenced by factors that vary day to day.

All these considerations motivate need for field tests. The seemingly inexorable growth in ride-sharing adds further motivation. Field tests would require certain accommodations. These could be met with careful thought, and the suitable application of technologies.

For example, more restrictive drop-off rules would mean that some patrons travel greater distances from their taxis to a station entrance. The onerousness of this might be lessened by providing luggage carts, human baggage carriers, and perhaps even moving walkways. Moreover, values of $L_0$ and $L_H$ might vary (e.g. over a day) with an eye towards inducing greater walk distances only when needed to battle taxi queueing upstream of the lane. Sensors (cameras, inductive loops, etc.) can be used to determine when changes in $L_0$ and $L_H$ might occur. Decision-making in this regard would improve with the acquisition of more data. Indeed, these data could eventually lead to all sorts of operational improvements. Sensors could likewise play a vital role in enforcing lane-use regulations.

**Acknowledgements**

The research was supported by a General Research Fund (No. 15217415) provided by the Research Grants Council of Hong Kong, and a project funded by National Science Foundation of China (No. 51178111). The authors thank Liang-peng Gao, Kui-sheng Xu, Qian Yu, Yang Yang, Hong-fei Hu, Mengmiao Liu, Rui Liu, Wan-yu Yang and Zhi-peng Liu of the School of Transportation at Southeast University (China) for their help with data collection.

**Appendix A. Trajectory point extraction method**

We present the method to convert the coordinates of a point on the video screen (the pixel coordinate system) to the coordinates of that point in the real world (the world coordinate system). The relationship between these two coordinate systems can be expressed as follows (Hartley and Zisserman, 2003):

$$S \begin{bmatrix} x \\ y \\ 1 \end{bmatrix} = A[R \quad T] \begin{bmatrix} X_W \\ Y_W \\ Z_W \\ 1 \end{bmatrix} \tag{A1}$$

where $S$ is an arbitrary scale factor, $(x, y)$ are the pixel coordinates of a point, and $(X_W, Y_W, Z_W)$ are the world coordinates of the same point.

Matrix $A$ is the camera intrinsic matrix determined by the camera's intrinsic properties, including focal length, image sensor format, principal point, radial distortion, etc. This matrix is given by:

$$A = \begin{bmatrix} f_x & 0 & x_0 \\ 0 & f_y & y_0 \\ 0 & 0 & 1 \end{bmatrix} \tag{A2}$$

where $x_0$ and $y_0$ are the coordinates of the principal point; and $f_x$ and $f_y$ are scale factors in the $x$ and $y$ axes in the pixel coordinate system, respectively (Hartley and Zisserman, 2003).

Matrix $[R \quad T]$ contains the external parameters, namely the camera's angle and position in the world coordinate system. Specifically, suppose the camera coordinate system is obtained from the world coordinate system by the following steps: rotate around the $X_W$ axis counterclockwise by $\theta$, around the $Y_W$ axis counterclockwise by $\psi$, and around the $Z_W$ axis counterclockwise by $\omega$; then translate by $T = [\Delta X, \Delta Y, \Delta Z]^T$. Then $R$ can be expressed by:

$$R = R_1(\theta) R_2(\psi) R_3(\omega) \tag{A3}$$

where,

$$R_1(\theta) = \begin{bmatrix} 1 & 0 & 0 \\ 0 & \cos\theta & \sin\theta \\ 0 & -\sin\theta & \cos\theta \end{bmatrix} \tag{A4}$$

$$R_2(\psi) = \begin{bmatrix} \cos\psi & 0 & -\sin\psi \\ 0 & 1 & 0 \\ \sin\psi & \cos\psi & 0 \end{bmatrix} \tag{A5}$$

$$R_3(\omega) = \begin{bmatrix} \cos\omega & \sin\omega & 0 \\ -\sin\omega & \cos\omega & 0 \\ 0 & 0 & 1 \end{bmatrix} \quad (A6)$$

We next calibrate the unknown parameters, including $x_0$, $y_0$, $f_x$, $f_y$, $\theta$, $\psi$, $\omega$, $\Delta X$, $\Delta Y$ and $\Delta Z$ (note the scale factor $S$ can be cancelled out after expanding (A1)). The calibration process takes two steps: step 1 for calibrating the intrinsic parameters $x_0$, $y_0$, $f_x$ and $f_y$, and step 2 for calibrating the external ones $\theta$, $\psi$, $\omega$, $\Delta X$, $\Delta Y$ and $\Delta Z$. Among the many calibration methods in the literature (e.g. Abdel-Aziz et al., 2015; Sturm and Maybank, 1999; Zhang, 2000; Tsai, 2003; Selby, 2011), we choose Zhang's method (Zhang, 2000) for step 1 due to its higher flexibility and robust estimates. This method works by using the camera to take pictures of a handy referencing object (with a planar pattern on its surface). The pixel coordinates $(x, y)$ and the world coordinates $(X_W, Y_W, Z_W)$ are obtained for a number of points on the referencing object's planar surface. (Note that the world coordinate system here is defined for the camera and the referencing object, not for the FIFO lane.) These coordinate data are then used to optimize both the intrinsic and external parameters via the maximum likelihood estimation (MLE) method. The intrinsic parameter matrix $A$ is obtained from the above step. The remaining work is to estimate the external parameter matrix $[R \quad T]$ for the world coordinate system where the same camera is used to picture the FIFO lane.

In our site, the origin of the world coordinate system is set at the vertical projection point of the camera on the road surface. The $Z_W$ axis is parallel to the taxi lane (pointing toward downstream); the $X_W$ axis is perpendicular to the taxi lane; and the $Y_W$ axis perpendicular to the road surface (thus $Y_W = 0$ for all the points on the road surface). In step 2, we use the conventional perspective projection transformation (Riley, 2002) to estimate $[R \quad T]$, because many other calibration methods (including Zhang's method) are not accurate for calibrating coordinates of far-range objects. To this end, we need to use at least three calibration points (whose pixel and world coordinates are known) to solve the six unknowns in matrix $[R \quad T]$, since each calibration point can produce two equations after expanding (A1) and eliminating $S$. In practice, we use dozens of calibration points to improve the accuracy. The approximate solution of the six unknowns to the resulting overdetermined system of equations is obtained via the ordinary least squares method.

After estimating $A$ and $[R \quad T]$, we test the error of (A1) by comparing the predicted locations ($Z_W$) of given points with their true locations measured on the site. Figure A1 illustrates the comparison for 32 test points located 0-80m from the camera location. The figure shows that the prediction error is very small but increases as the test point is placed farther from the camera. This is why in our method each camera was used to cover a range of 70m or less.

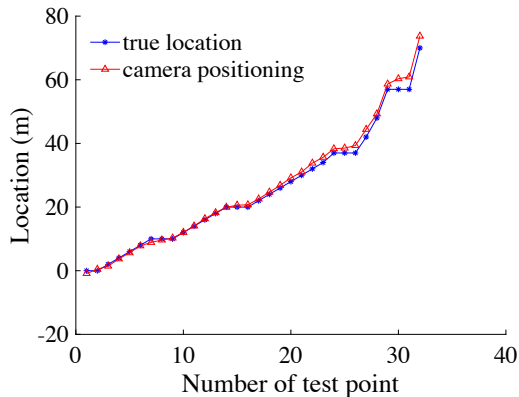

Figure A1. Error test of the trajectory point extraction method

## Appendix B. The *k*-mean clustering method

The FIFO lane was partitioned by clustering taxis' forced wait times at their first instances of forced stops. Taxis' second, third and fourth instances of forced stops were excluded because they were of much shorter durations. For a given number of segments $k$, we seek a partition that minimizes the sum of total squared errors of forced wait times in each segment, $\varepsilon$:

$$\min_{C \triangleq \{C_1, C_2, \ldots, C_k\}} \varepsilon = \sum_{i=1}^{k} \sum_{n: y^n \in C_i} (t^n - u_i)^2, \tag{B1}$$

where $C$ is a lane partition, with each $C_i$ ($i = 1, 2, \ldots, k$) defining a continuous space interval (i.e. a segment) in $[0, 240m]$, $\bigcup_{i=1}^{k} C_i = [0, 240m]$; $y^n$ is the location of taxi $n$'s first forced stop (given that the taxi is not leading a batch); $t^n$ is taxi $n$'s forced wait time during that stop; and $u_i = E[t^n | y^n \in C_i]$.

## Appendix C. Estimation of patience distribution

The probability density function (PDF) of a mixture distribution for patrons' patience is given as:

$$f(p) = \gamma f_1(p) + (1 - \gamma) f_2(p), \tag{C1}$$

where $f_1(p)$ and $f_2(p)$ are the PDFs of patience distributions for impatient patrons (i.e., those who alighted almost immediately after being forced to stop) and the remaining, patient ones, respectively; and $\gamma$ is the probability that a taxi's patron(s) were impatient. Specifically, when $f_1(p)$ and $f_2(p)$ are gamma PDFs, we have:

$$f(p; \gamma, k_1, \theta_1, k_2, \theta_2) = \gamma f_1(p; k_1, \theta_1) + (1 - \gamma) f_2(p; k_2, \theta_2)$$
$$= \gamma \frac{p^{k_1 - 1} e^{-p/\theta_1}}{\theta_1^{k_1} \Gamma(k_1)} + (1 - \gamma) \frac{p^{k_2 - 1} e^{-p/\theta_2}}{\theta_2^{k_2} \Gamma(k_2)}, \tag{C2}$$

where $k_1$ and $k_2$ are the shape parameters, and $\theta_1$ and $\theta_2$ are the scale parameters for $f_1(p)$ and $f_2(p)$, respectively; and $\Gamma(\cdot)$ is the gamma function.

To estimate the values of $k_1$, $k_2$, $\theta_1$, $\theta_2$ and $\gamma$, we formulate the log-likelihood function for the forced wait times as:

$$LL(\gamma, k_1, \theta_1, k_2, \theta_2) = \sum_{n \in \mathcal{P}} \ln f(t^n, \gamma, k_1, \theta_1, k_2, \theta_2) + \sum_{n \in Q} \ln[1 - F(t^n, \gamma, k_1, \theta_1, k_2, \theta_2)], \tag{C3}$$

where $\mathcal{P}$ denotes the index set of taxis that dropped-off patrons at the present forced stop (i.e., the taxis whose forced waits equaled their patience); $Q$ denotes the index set of taxis that did not drop-off patrons at the forced stop (i.e., those whose forced waits were less than their patience); and $F(\cdot)$ is the CDF of the mixture distribution.

The MLE problem is then formulated as:

$$\max_{\gamma, k_1, \theta_1, k_2, \theta_2} LL. \tag{C4}$$

This problem was solved by the nonlinear program solver "fminsearch" in Matlab R2017b.

The distribution parameter estimates for the two datasets are presented in Tables C1 and C2. Note in each day's data that, for all the five distributions, the mean and variance for the impatient patrons are much smaller than those for the remaining, patient patrons; i.e., $k_1\theta_1 \ll k_2\theta_2$ and $k_1\theta_1^2 \ll k_2\theta_2^2$ for all the five rows of each table. Also, both tables show that the probability of impatient patrons, $\gamma$, increases from Segment 1 to Segment 4. This is consistent with intuition, since patrons became less patient as they moved downstream.

Table C1. Optimal parameters for patience distribution (the April 25 dataset)

|  | $k_1$ | $\theta_1$ | $k_1\theta_1$ | $k_1\theta_1^2$ | $k_2$ | $\theta_2$ | $k_2\theta_2$ | $k_2\theta_2^2$ | $\gamma$ |
|---|---|---|---|---|---|---|---|---|---|
| 1st-instance forced stops in Segment 1 | 2.13 | 1.42 | 3.02 | 4.29 | 3.62 | 8.77 | 31.8 | 278.4 | 0.43 |
| 1st-instance forced stops in Segment 2 | 1.81 | 1.25 | 2.26 | 2.83 | 3.63 | 7.29 | 26.5 | 192.9 | 0.48 |
| 1st-instance forced stops in Segment 3 | 0.97 | 6.80 | 6.60 | 44.85 | 5.71 | 4.78 | 27.3 | 130.5 | 0.55 |
| 1st-instance forced stops in Segment 4 | 1.10 | 2.15 | 2.37 | 5.08 | 6.25 | 3.17 | 19.8 | 62.8 | 0.64 |
| 2nd~4th-instance forced stops | 3.48 | 0.70 | 2.44 | 1.71 | 3.75 | 8.72 | 32.7 | 285.1 | 0.49 |

Table C2. Optimal parameters for patience distribution (the July 13 dataset)

|  | $k_1$ | $\theta_1$ | $k_1\theta_1$ | $k_1\theta_1^2$ | $k_2$ | $\theta_2$ | $k_2\theta_2$ | $k_2\theta_2^2$ | $\gamma$ |
|---|---|---|---|---|---|---|---|---|---|
| 1st-instance forced stops in Segment 1 | 17.09 | 0.17 | 2.91 | 0.49 | 2.71 | 14.74 | 39.95 | 588.80 | 0.25 |
| 1st-instance forced stops in Segment 2 | 14.67 | 0.16 | 2.35 | 0.38 | 1.84 | 14.57 | 26.81 | 390.60 | 0.40 |
| 1st-instance forced stops in Segment 3 | 3.92 | 1.05 | 4.12 | 4.32 | 4.44 | 7.23 | 32.10 | 232.09 | 0.36 |
| 1st-instance forced stops in Segment 4 | 6.27 | 0.47 | 2.95 | 1.39 | 5.93 | 4.58 | 27.16 | 124.39 | 0.56 |
| 2nd~4th-instance forced stops | 20.02 | 0.13 | 2.60 | 0.34 | 1.70 | 13.01 | 22.12 | 287.74 | 0.31 |